\documentclass[12pt]{article}
\usepackage{amssymb}
\usepackage{cite}
\newlength{\abstractwidth}
\usepackage[dvips]{graphicx}
\usepackage{subfigure}

\renewcommand{\thefootnote}{\fnsymbol{footnote}}
\renewcommand{\thanks}[1]{\footnote{#1}} % Use this for footnotes
 \newcommand{\starttext}{
\setcounter{footnote}{0}
\renewcommand{\thefootnote}{\arabic{footnote}}}
\renewcommand{\theequation}{\thesection.\arabic{equation}}
\newcommand{\be}{\begin{equation}}
\newcommand{\bea}{\begin{eqnarray}}
\newcommand{\eea}{\end{eqnarray}}
\newcommand{\beq}{\begin{equation}}
\newcommand{\ee}{\end{equation}}
\newcommand{\eeq}{\end{equation}}

\newcommand{\<}{\langle}

\newcommand{\comment}[1]{}
\renewcommand{\>}{\rangle}
\def\ba{\begin{eqnarray}}
\def\ea{\end{eqnarray}}

\def\t0{$t=0$}
\def\14{{1\over4}}
\def\12{{1 \over 2}}

\def\h3{h^{3\over 2}}

\def\>{\rangle}
\def\<{\langle}

\def\0cc{$\Lambda = 0$}

\def\om{   \Omega_{D-2}^2   }

\begin{document}
\renewcommand{\theequation}{\thesection.\arabic{equation}}
\begin{titlepage}
\bigskip
\rightline{UCB-PTH-05/33} \rightline{LBNL-58948} 
\rightline{hep-th/0511084}

\bigskip\bigskip\bigskip\bigskip

\centerline{\Large \bf {Asymptotic states of the bounce geometry}}

\bigskip\bigskip
\bigskip\bigskip

\centerline{\bf Raphael Bousso and Ben Freivogel}
\medskip
\centerline{\small Department of Physics and Center for Theoretical
Physics} 
\centerline{\small 
University of California, Berkeley, California 94720, U.S.A.}
\medskip
\centerline{\small Lawrence Berkeley National Laboratory, 
Berkeley, California 94720, U.S.A.}
\medskip
\medskip

\bigskip\bigskip \begin{abstract}
We consider the question of asymptotic observables in cosmology.  We
assume that string theory contains a landscape of vacua, and that
metastable de~Sitter regions can decay to zero cosmological constant
by bubble nucleation.  The asymptotic properties of the corresponding
bounce solution should be incorporated in a nonperturbative quantum
theory of cosmology.  A recent proposal for such a framework defines
an S-matrix between the past and future boundaries of the bounce.  We
analyze in detail the properties of asymptotic states in this
proposal, finding that generic small perturbations of the initial
state cause a global crunch. We conclude that late-time amplitudes
should be computed directly. This would require a string theory
analogue of the no-boundary proposal.
\end{abstract}

\end{titlepage}
\starttext \baselineskip=18pt \setcounter{footnote}{0}

%%%%%%%%%%%%%%%%%%%%%%%%%%%%%%%%%%%%%%%%%%%%%%%%%%%%%%%%%%%%%%%%%%%%%%
%%%%%%
%%%%%%%%%%%%%%%%
%%%%%%%%%%%%%%%%%%%%%%%%%%%%%%%%%%%%%%%%%%%%%%%%%%%%%%%%%%%%%%%%%%%%%%
%%%%%%
%%%%%%%%%%%%%%%%

\setcounter{equation}{0}

\section{Introduction}

The observed acceleration of the expansion of the universe has
modified the cosmological constant problem.  In Ref.~\cite{BP} it was
argued that string theory contains an exponentially large number of
long-lived false vacua, all of which are realized dynamically as huge
regions in an infinite universe, and some of which have an effective
cosmological constant of order the observed value.  Recent progress on
stabilizing string compactifications%
~\cite{KKLT,DasRaj99,GidKac01,DenDou05,DenDou04,ConQue05,%
  BerMay05,BalBer05,SalSil04,SalSil204,MalSil02,Ach02,CarLuk04,%
  CurKra05,GurLuk04,BecBec03,BecCur04,BruAlw04,GukKac03,%
  BucOvr03,DewGir05} %
lends support to this scenario and provides increasingly concrete
explorations of the stringy ``landscape''~\cite{Sus03}.

Another, more conceptual problem is the question of observables in
cosmology.  A number of authors have pointed out that the definition
of exact observables is problematic in cosmology (see, e.g.,
Refs.~\cite{BanFis01a,Wit01,HelKal01,KraOog03,Ban04,Bou05,FreHub05})
because of obstructions such as event horizons, thermal radiation, and
entropy bounds.  In particular, in Ref.~\cite{Bou05} it was argued
that an S-matrix cannot be measured by any real observer inside a
large class of universes.  However, it was also noted that event
horizons do not automatically preclude the existence of other exact
observables.  In any case, if exact observables can be defined at all,
they will be defined at asymptotically late time.

There is no reason why a rigorous quantum gravity theory with
well-defined observables should exist for all imaginable cosmologies;
indeed, it would be satisfying if its existence were to provide a
criterion constraining the type of universe we find ourselves
in. Assuming that the landscape picture is
correct,\footnote{Disclaimers should include the fact that corrections
  to metastable de~Sitter solutions are not yet controlled with the
  same level of rigor as in supersymmetric backgrounds.  Because of
  our potential ignorance of vast sectors of string theory, current
  surveys of vacua may not be representative.  For further criticism,
  see, e.g., Ref.~\cite{BanDin03}.} it makes sense to search for a
nonperturbative theoretical framework that rigorously defines its
dynamics and observables.

A first step in this direction was taken in Ref.~\cite{FreSus04},
which noted that the asymptotic structure of supersymmetric regions
with vanishing cosmological constant may allow for constructions of
observables that would be impossible in an eternal de~Sitter universe.
(Regions with negative vacuum energy do not share this property
because they collapse when accessed cosmologically.)  The
corresponding semiclassical solution is the Coleman-De~Luccia
``bounce''.  It consists of a bubble of a $\Lambda = 0$ vacuum inside
de~Sitter space. The bubble of true vacuum expands but the de~Sitter
region inflates fast enough that it is not consumed. Asymptotically at
early and late times, the geometry has infinite regions of true and
false vacuum.  It was suggested that this system can be described
quantum mechanically in terms of an S-matrix relating asymptotic
states in the early and late time $\Lambda =0$ regions.

In the present paper we present results that suggest a partial
modification of this strategy.  In Sec.~\ref{sec-bound} we identify an
entropy bound limiting the number of states in the semiclassical
geometry expected to dominate the path integral.  This is concrete
evidence that the proposed S-matrix would have finite rank, as first
suggested by Banks~\cite{Ban04}.  We point out that the past half of
the Coleman-De~Luccia evolution violates the second law of
thermodynamics.  We find that this makes it difficult to characterize
the asymptotic states relevant to the proposed S-matrix.

In Sec.~\ref{sec-instability} we analyze how the entropy bound is
reconciled at the semiclassical level with the infinite number of
particles present in the universe at early and late times.  Here we
find that the picture is more complex than previously
noted~\cite{Ban04}.  By construction, the time-symmetric
Coleman-De~Luccia solution involves an era in which the coarse grained
entropy decreases.  This leads to violent instabilities.  Even a small
perturbation in the initial state will generically produce a big
crunch rather than a future asymptotic region.  In this respect the
Coleman-De~Luccia geometry considered here is not better, but worse
than eternal de~Sitter space.

In Sec.~\ref{sec-halo} we characterize ``allowed'' states that
correspond to a semiclassical evolution from a past to a future
asymptotic region.  Many such states do not appear to admit a simple
description near the boundary at all.  Those that do are best thought
of as excitations of zero mass in a cosmological fluid.  The task is
to leave all fields outside a compact region undisturbed, in order to
limit the crunch to a finite black hole.  We achieve this by treating
all matter as dust and creating an overdense region surrounded by an
underdense region (a ``halo'').

Working with such states is awkward, and the relation between their
S-matrix elements and bulk observables accessible to physical
observers is obscure.  In light of the inhomogeneity of the allowed
perturbations we found, the corresponding S-matrix is unlikely to
describe the perturbations that are actually observed in the universe
at late times.  Unless the dust approximation is exact, even halo
states involve infinitely fine-tuned small adjustments in the entire,
infinite initial $\Lambda=0$ region.  Moreover, the choice of initial
state has no natural interpretation and may be redundant.  Note that
all of these difficulties relate to the {\em past\/} asymptotic
region.

Yet, in light of the difficulty in defining observables in cosmology,
we would like to exploit the more advantageous asymptotic properties
of the Coleman-De~Luccia solution.  They are associated with the {\em
  future\/} asymptotic region.  It is particularly promising that the
solution contains an open (rather than closed or flat) FRW universe,
since this cuts off the growth of perturbations after curvature
domination. Arbitrarily large regions of arbitrarily flat space become
available at late times.  In particular, there will be well-separated,
noninteracting particles. Just as string theory computes the S-matrix
in flat space, it may well compute amplitudes for out-states in the
Coleman-De~Luccia geometry.

Thus, our results suggest that the second-law-violating, contracting
half of the Coleman-De~Luccia universe does more to obstruct than to
facilitate a nonperturbative description of the landscape and its
dynamics.  They point instead to a framework in which amplitudes are
computed for boundary conditions specified {\em at late times only}.
An example of this in semiclassical gravity is the Hartle-Hawking
wavefunction~\cite{HarHaw83}, defined by the saddlepoint approximation
to a path integral over compact geometries.  In the context of string
theory, however, one would seek an exact nonperturbative definition of
amplitudes in the asymptotic future.

\section{Metric and causal structure}
\label{sec-metric}

In this section we review the fully extended Coleman-De~Luccia
solution~\cite{ColDel80}, or ``bounce''.  We present the metric and
conformal diagram.  More details can be found, e.g., in
Refs.~\cite{Ban02,FreSus04,Bou05}. We work in four dimensions for
simplicity, but our analysis is independent of dimension aside from
trivial factors.

\begin{figure}[!htb]
\center
\includegraphics [scale=.5, clip]
{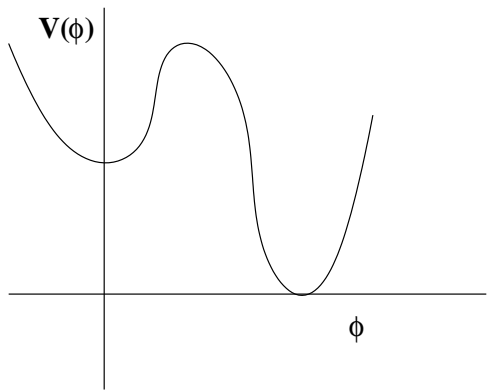} \caption{A potential which has a metastable
  minimum with positive energy and a stable minimum with
  zero energy.}
\label{fig-pot}
\end{figure}

The bounce is a solution to the Einstein equations in the presence of
a scalar field potential with at least one false vacuum, as shown in
Fig.~\ref{fig-pot}.  We assume that the cosmological constant vanishes
exactly in the true vacuum.  The simplest classical solutions are
Minkowski space (where the scalar is everywhere in the true vacuum)
and de~Sitter space (with the scalar in the false vacuum).  Both of
these have maximal symmetry, generated respectively by the Poincar\'e
group and the de~Sitter group, $SO(1,4)$.  The de~Sitter curvature
radius $R$ is inversely related to the cosmological constant $\Lambda$
in the false vacuum, $R= \sqrt{3/\Lambda}$.

But there is also a field configuration in which the scalar crosses
the barrier from true to false vacuum, forming a domain wall.  The
bounce is the most symmetric solution of the Einstein equation with
this field configuration.  Its isometry group is $SO(1,3)$. For
simplicity, we will describe the solution in the thin wall limit where
the width of the bubble is negligible compared to its minimum radius.
(See Ref.~\cite{ColDel80} for the conditions on the potential under
which this limit is valid.)

Inside the domain wall, the spacetime is approximately empty, so the
geometry is simply a region of Minkowski space glued across the thin
domain wall to a portion of de Sitter space.  This approximation
actually requires infinite fine tuning, so it will important to
include the effects of matter present in the $\Lambda=0$ region.
However, it is instructive to describe first the geometry of the
idealized, vacuous bounce.  It will then not be difficult to go beyond
this excessive simplification.

The domain wall is a hyperboloid, whose worldvolume describes a
three-dimensional de Sitter spacetime.  Its minimum area is $4\pi
r_0^2$, where the parameter $r_0$ is determined by the tension
$\sigma$ of the domain wall and the ($4$-dimensional) de Sitter radius
$R$, 
\begin{equation} 
r_0 = {8 \pi G \sigma R^2 \over (4 \pi G
  \sigma R)^2 + 1 }~.  
\end{equation}

\begin{figure}[!htb]
\center
\includegraphics [scale=.5]
{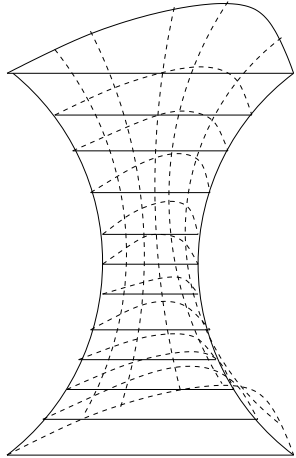} \caption{The Coleman-De~Luccia bounce geometry,
  as an embedding in Minkowski space.  The flat piece has zero cosmological
  constant, while the curved piece has positive cosmological constant.
  A domain wall separates the two regions.} \label{embd}
\end{figure}

In the flat spacetime enclosed by the hyperboloid the position of the
domain wall satisfies
\be r^2 - t^2 = r_0^2~, \ee
where $r$ and $t$ are the standard Minkowski coordinates.  The wall
expands from a minimum radius $r_0$, at $t=0$, to infinite size as
$t\to\pm\infty$.  To describe the motion of the domain wall from the
de Sitter side, it is simplest to embed de Sitter space in
$5$-dimensional Minkowski space as the surface
\be X_\mu X^\mu = -R^2, ~~~~~\mu = 0,1,...,4~. \ee
The domain wall is the intersection of the de Sitter space with
the plane
\be X_4 = (R^2 - r_0^2)^{1/2}~. \ee
The part of the space farther from the origin than the plane is
discarded and replaced by a piece of flat space. An embedding
picture of the solution is shown in Fig.~\ref{embd}.

\begin{figure}[!!htb] \center \includegraphics [scale=.8, clip]
      {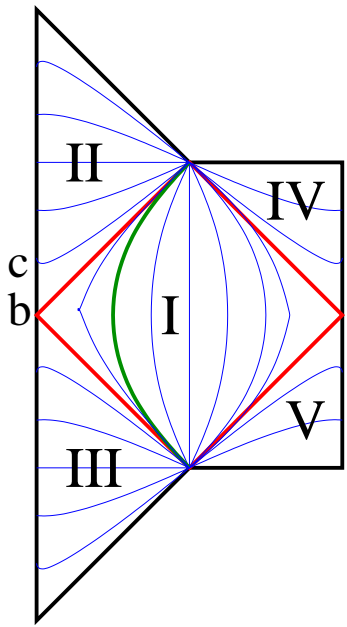} \caption{Conformal diagram of the
        Coleman-De~Luccia solution, including orbits of the symmetry
        group $SO(1,3)$. The thick orbit is the domain wall.  The
        radius of spheres goes to zero at the left and right
        boundaries and to infinity on the remaining boundaries.}
      \label{fig-benpen} \end{figure}

The solution is invariant under the symmetry group of the hyperboloid,
$SO(1, 3)$. On the de Sitter side, the full $SO(1,4)$ de Sitter
symmetry is broken down to those boosts and rotations which leave
$X_4$ unchanged. On the Minkowski side, the symmetry arises because
the domain wall picks out an origin, breaking translation invariance
while preserving the Lorentz group. The orbits of the symmetry group
are surfaces of constant invariant distance from the origin. These
surfaces can be spacelike separated from the origin, like the domain
wall (region I of Fig.~\ref{fig-benpen}). They can also be timelike
separated from the origin (region II of Fig.~\ref{fig-benpen}).
The surviving $SO(1, 3)$ symmetry is not an artifact of the vacuous
bounce.  It is present in the full solution (even beyond the thin wall
approximation). 

If the thin wall approximation is well satisfied, the energy density
in the center of the true vacuum bubble (point $b$ in
Fig.~\ref{fig-benpen}) will almost vanish.  But it will not be exactly
zero, because the distance of $b$ to the domain wall is finite, so
that $b$ lies on the exponential tail along which $\phi$ attempts to
approach the true vacuum.  By continuity, the same nonzero energy
density will be present on the infinitesimally later point $c$, and
hence {\em on the entire symmetry orbit of c}.  The presence of a
constant energy density on a hyperbolic spacelike slice means that
region II (III) is not flat space, but an expanding (contracting) open
FRW cosmology~\cite{ColDel80}, with metric 
\begin{equation} 
ds^2 = -d\tau^2 + a^2(\tau) [d\rho^2 + \sinh^2 \rho d\om].  
\label{eq-iop}
\end{equation} 
Thus, the idealization as flat spacetime is not even approximately
correct in regions II and III, in the sense that the total energy in
those regions is infinite.

As the field $\phi$ completes its descent into the true vacuum, it
will need to dissipate this extra energy.  For simplicity, we think of
the conversion of potential energy to particles (``reheating'') as
occurring on a definite time slice of the FRW geometry, the reheating
surface.  By coupling $\phi$ to other fields, particles are produced
at the reheating surface. The details can vary, but it is inevitable
that somehow the initially uniform potential energy will evolve into
an incoherent, highly entropic form as the universe expands. Because
the symmetry orbits are noncompact hyperboloids, an infinite number of
particles will be produced.

\section{Entropy}
\label{sec-bound}

The bounce solution has an infinite amount of matter, and hence, an
infinite amount of coarse-grained entropy in the FRW regions.  In this
section, we apply the covariant entropy bound~\cite{CEB1,CEB2} to show
that the total number of allowed microstates is nevertheless finite.

A light-sheet is a convergent null hypersurface orthogonal to an
arbitrary two-dimensional spatial surface with area $A$.  The bound
says that there can be at most $e^{A/4G}$ different quantum states on
a light-sheet, i.e., the entropy will not exceed $A/4G$ (see
Ref.~\cite{RMP} for a detailed review).
\begin{figure}[!htb] \center \includegraphics [scale=.6] {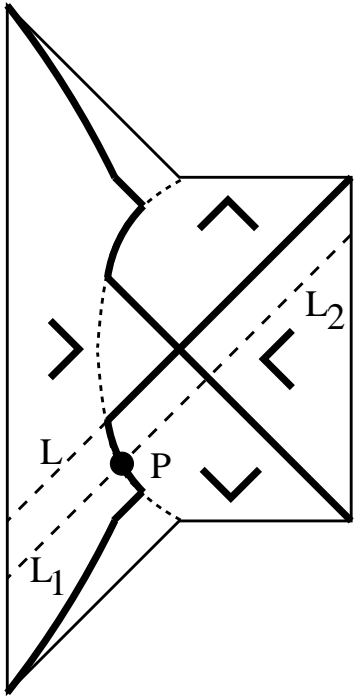}
\caption{A wedged conformal diagram~\cite{CEB2} for the bounce
  geometry.  Normal ($>$, $<$), trapped ($\vee$), and anti-trapped
  ($\wedge$) regions are separated by apparent horizons (thick lines).
  Their shape is determined by matter sources (the domain wall, and
  the matter created by it in regions II and III).  We treat the
  domain wall (dotted line) as a delta function source.  $L_1$ and
  $L_2$ are light-sheets of the sphere $P$.  The tightest entropy
  bound on the spacetime is obtained from $L$, a single light-sheet
  whose maximal area is the de~Sitter horizon in the top right
  corner.}  \label{fig-cdlbp} \end{figure}

In a spherically symmetric spacetime, each sphere can be classified
depending on the contracting null directions, i.e., the directions in
which light-sheets exist.  The corresponding domains are indicated in
Fig.~\ref{fig-cdlbp}: spheres are either trapped ($\vee$), antitrapped
($\wedge$), or normal ($>$ and $<$, with the open side pointing
towards $r=0$).

As is common for a cosmological solution, the bounce geometry contains
light-cones of finite maximal area.  By following the wedges back to
the tips, one is led to an apparent horizon, an area maximal with
respect to orthogonal light-rays.  Such a surface, for example the
sphere $P$ in Fig.~\ref{fig-cdlbp}, admits light-sheets in two
opposing directions.  These two light-sheets, $L_1$ and $L_2$, form a
complete Cauchy surface, in the sense that every timelike curve must
pass through them.  Hence the total number of states in the universe
is bounded by $e^{A_P/2G}$.

Note that $P$ is a sphere on the worldvolume of the domain wall.  Its
area increases monotonically and without bound as $P$ is moved down.
This explains why we can have an infinite coarse-grained entropy at
early times in the FRW region.  We get the best possible bound by
moving $P$ all the way up until the light-sheet becomes $L$.  This is
a {\em single\/} light-sheet starting at the late-time cosmological
horizon of the de~Sitter region (the top right corner).  Hence the
optimal bound on the total number of states is given by
\begin{equation}
N \leq \exp\left(\frac{\pi R^2}{G} \right)~,
\label{eq-bound}
\end{equation}
where $R$ is the de~Sitter radius.  (One might worry about what
happened to the second light-sheet: is this estimate not off by a
factor of 2?  There is indeed a de~Sitter horizon volume in the top
right corner which is not captured by $L$.  However, the
D-bound~\cite{Bou00b} implies that no matter entropy is present there.)

Thus there is a huge discrepancy: Unbounded coarse-grained entropy is
allowed, and indeed present, at early and late times in the FRW
regions II and III; yet, the total number of microstates allowed to
pass though the light-sheet $L$ in the center of the geometry is
finite.  It follows that of the infinite number of microstates
corresponding to the initial macroscopic configuration, all but a
final number will not reach the light-sheet $L$.  What enforces this
pruning dynamically?

\section{Instabilities}
\label{sec-instability}

In this section we discuss why many initial states that are
macroscopically identical to that of the background lead to a
completely different future evolution.  It is instructive to consider
first the idealized case where no energy density is present in the
true vacuum region of the background.  This case turns out to be
somewhat similar to de~Sitter space~\cite{Ban00,Bou00a}: the
background is stable against sufficiently small perturbations, but
introducing too much matter will make it collapse to a big
crunch.  The extra mass produced near the time-symmetric slice is a
good diagnostic for the permissibility of a perturbation.

Interestingly, this is no longer the case for the true bounce
solution, which is filled with homogeneous matter energy density.  We
find that this background is far more unstable.  The process whereby
matter energy is absorbed into vacuum energy, permitting a bounce, is
extremely delicate since it decreases coarse-grained entropy.
Arbitrarily small perturbations at early times typically suffice to
derail this absorption completely, causing a big crunch.  (On the
other hand, one might imagine introducing enormous extra mass at early
times; this could be consistent with a bounce as long as it all ends
up in a coherent excitation of the field $\phi$.)

\subsection{Instabilities of the vacuous bounce}
\label{sec-babyinst}

In this subsection we assume that by fine tuning the
potential, we have arranged that there are no particles in the
background.  In this case the matter entropy vanishes at all times.

In simple examples, it is possible to see how gravitational
backreaction enforces the bound (\ref{eq-bound}). For example, we
could add a black hole at rest at the center of the flat region. We
will see that there is a bound on how big the black hole can be.

It is simplest to analyze the situation on the time-symmetric slice
\t0; because we are adding a black hole at rest at the center it does
not break the time reversal invariance of the background.  The
solution will be Schwarzschild out to the domain wall. Outside the
domain wall, the geometry must be de Sitter or Schwarzschild-de Sitter
because of the spherical symmetry. The radius of the domain wall,
$r_0$, will be bigger~\cite{FreSus04} than the Schwarzschild radius of
the mass, $2GM$. In order to match to de Sitter space along the
time-symmetric slice, the domain wall radius must be less than the de
Sitter radius $R$. In order to satisfy both of these constraints, we
need $2GM < R$. A bound on the mass of the black hole is equivalent to
a limit on the entropy, \be S < {\pi R^2 \over G}, \ee which agrees
precisely with the prediction we derived from the covariant entropy
bound.

Another example is to add $N$ photons to the initial state.  Because
the spatial slices are infinite, their wavelength can be made
arbitrarily long.  Hence $N$ can be large without causing large energy
density.  The entropy of the state is $N$, and if the entropy exceeds
the entropy bound (\ref{eq-bound}) then something must go wrong. What
happens is that as the space contracts towards $t=0$, the photons are
blueshifted. At $t=0$, the wavelength of the photons is limited by the
size of the space, roughly the de Sitter radius $R$. $N$ photons with
wavelength $R$ have energy $N/R$.  If $N$ exceeds the entropy bound
$S_{max}$ then the energy exceeds roughly $R/G$.

There must be a huge backreaction, because the biggest black hole that
can fit in de Sitter space has a mass of order
\begin{equation}
M_{\rm BH}\approx R/G~.
\end{equation}
If we exceed the entropy bound, we are inserting more energy than can
fit through the geometry, even if we allow black hole formation. The
result must be a big crunch.
Thus we find again that the conditions for avoiding a crunch and for
avoiding a violation of the bound coincide.

\begin{figure}[!!htb] \center \includegraphics [scale=.7, clip]
      {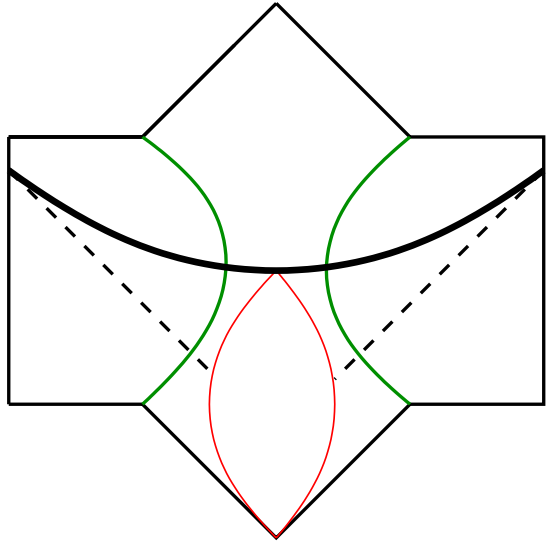} \caption{Two ordinary particles collide with a
        large center of mass energy if they begin far apart. The
        backreaction will cause a crunch (heavy line).  The regions
        above this line are included only to guide the eye.  The
        diagram has been doubled: every point represents a
        hemisphere, not a sphere.}  \label{collision} \end{figure}

Of course, not all states that lead to a crunch have large entropy.
Consider adding two small particles of mass $m$ which are comoving
with respect to the open-FRW coordinate system picked out by the
symmetries of the geometry.  As we shall see, they can be more simply
characterized from the point of view of Minkowski space, as two
particles colliding at the event $r=0$, $t=0$, as shown in
Fig.~\ref{collision}.
The FRW coordinates cover a region of flat space, with metric
\be
ds^2 = -d\tau^2 + \tau^2(d\rho^2 + \sinh^2\rho\, d\om)~.
\ee
We use Greek letters $(\tau,\rho)$ for the FRW coordinates and Roman
letters $(r,t)$ for the usual flat space coordinates. The two
coordinate systems are related by
\bea
\tau^2 & = & t^2 - r^2 \\
\sinh^2\rho & = & {r^2 \over t^2 - r^2}~.
\eea
A particle at rest in FRW coordinates
follows the trajectory
\begin{equation}
r(t) = t \tanh \rho_0~,
\end{equation}
where $\rho_0$ is the value of its FRW radial coordinate. The
trajectory passes through the origin at $t=0$ no matter where in the
FRW the particle began. So two particles at rest relative to the FRW
coordinates will collide at the origin. If they start at opposite
points on the 2-sphere, their relative velocity will be
\begin{equation}
v_{\rm rel} = 2 \tanh \rho_0 ~.
\end{equation}
The center of mass energy is
\begin{equation}
E_{\rm cm} = 2 m \cosh \rho_0~,
\end{equation}
where $m$ is the mass of the particles, so the energy is large if they
start far apart.  To avoid a global crunch, this energy must not
exceed the mass of the maximal black hole, leading to the
inequality
\begin{equation}
m \cosh \rho_0 \lesssim R/G~.
\end{equation}
This result is interesting because it indicates that two innocent
perturbations far from each other tend to collide catastrophically.

\subsection{Instabilities of the bounce}
\label{sec-fullinst}

As discussed in Sec.~\ref{sec-metric}, the FRW regions (II and III) of
the bounce contain an infinite number of particles. The light-sheet
$L$ can contain only a finite number of particles, so as time evolves,
all but a finite number of particles must be absorbed. For
definiteness, we assume that in the background all particles are
produced on a reheating surface in region II, so by time-reversal
symmetry, they must be absorbed on a ``recooling'' surface in region
III, prior to $t=0$.

\begin{figure}[!htb] \center \includegraphics [scale=.4]
      {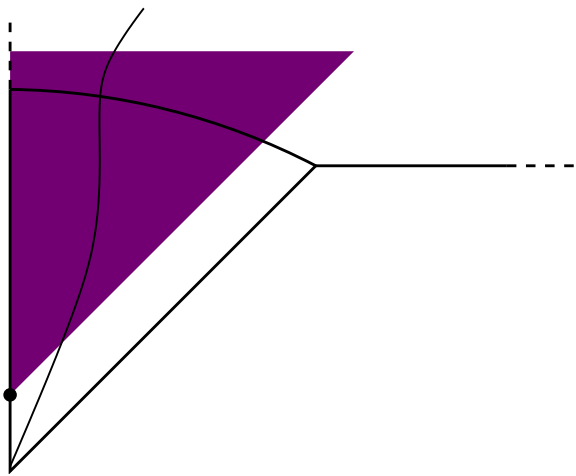} \caption{The past half of the bounce solution.  A
        perturbation (dot) will affect everything in its future
        lightcone (shaded), in general destroying the delicate
        ``recooling" process (at the heavy line).  Comparison to the
        apparent horizon (wiggly line) shows that this can cause a
        large number of horizon volumes to remain matter-dominated.}
      \label{unstable} \end{figure}

Recooling is an extremely delicate process that decreases the entropy
drastically.  It is comparable to a broken glass reassemling
spontaneously, or all air molecules collecting in one small corner of
a room.  Only a miniscule subset of phase space trajectories
correspond to these processes.  Tiny perturbations do not take these
special states into each other, but into states with a generic
evolution.  For example, a tiny mass added in the far past will
gravitationally influence a vast number of other particles and disturb
the recooling process in the entire future light-cone of the
perturbation (see Fig.~\ref{unstable}).  If the perturbation is
introduced sufficiently early, this will result in an enormous number
of particles remaining unabsorbed.  {\em Their\/} energy, not the
energy of the initial perturbation, decides whether the universe will
bounce or crunch.

In fact, recooling can be thrown off just by moving some particles
around, leaving the energy invariant (at least at the perturbative
level at which we might hope to define energy in this background); or
by removing some particles from the background, effectively decreasing
the initial energy and entropy.

But the entropy bound does not forbid all states; there should be a
finite number that do not destroy the geometry.  In the next section,
we construct examples of such states and verify that their entropy
approximately saturates the bound.

\section{Allowed perturbations}
\label{sec-halo}

In the saddlepoint approximation to the path integral, amplitudes
between points near past and future infinity should correspond to
non-crunching perturbations.  In this section we provide approximate,
semiclassical constructions of two classes of such states.

\subsection{States defined at \t0}

A natural place to define amplitudes is the past and future FRW
infinities.  But we have seen in the previous section that it is
tricky to identify in-states that do not collapse.  At least at the
semi-classical level, however, such states can be found by perturbing
the geometry on the time-symmetric slice, $t=0$.  (In a full quantum
theory, amplitudes would receive contributions from various
configurations at $t=0$.)

Unlike the FRW regions, the true vacuum portion of the time-symmetric
slice is well approximated by empty flat space.  Hence we can use the
results of Sec.~\ref{sec-babyinst}.  There are small perturbations
with a few extra particles, but most states will correspond to black
holes.  As discussed in Sec.~\ref{sec-babyinst}, the largest black
hole that fits has radius of order $R$, so the entropy of the possible
perturbations will be of order $R^2/G$, in agreement with the bound
(\ref{eq-bound}).

When those states propagate forward, they undergo complex interactions
with the particles produced at reheating.  Some of the excitations
will cross the event horizon of the FRW observer and enter region
IV. In Ref.~\cite{FreSus04} it was conjectured that the evolution is
nevertheless unitary.  But if scattering off of black holes offers any
guidance, the information will arrive at ${\cal I}^+$ in extremely
scrambled form.

Thus, it will be hard to look at an out-state and tell whether it
originated from an allowed perturbation at $t=0$, i.e., whether it
corresponds semiclassically to a geometry with that contains also a
past asymptotic region.  But by time-reversal symmetry of the bounce
background, this means it is also very difficult to set up an initial
perturbation that will not crunch, but will rather bounce and produce
a late-time geometry asymptotic to the future regions of the bounce.

In the next subsection, we shall see that it is nevertheless possible
to characterize some of the suitable in-states approximately, and to
verify once more that their entropy is in agreement with the bound
(\ref{eq-bound}).

\subsection{Halo states}

We are interested in an approximate construction, near ${\cal I}^-$,
of asymptotic states that can lead to a nonsingular bounce geometry.
We will assume that the matter present in the FRW regions is dust.  We
will also assume that its density is low enough so that the FRW
evolution is always curvature dominated.

The basic idea is to construct solutions which differ from the bounce
in a finite region, so that the delicate ``recooling'' process is only
disturbed for a small number of particles.  If we could neglect
gravitational signals propagating out from the perturbation, this
would be readily accomplished by adding, subtracting, or rearranging
particles in a bounded region of space.  But we must be careful to
avoid any changes, no matter how small, in the gravitational field
outside the region considered.

We achieve this by imposing two conditions.  The first is that the
perturbation be spherically symmetric, so that Birkhoff's theorem
applies.  The second is that the total mass in the perturbed region
remain unchanged compared to the background.  A simple example is an
overdense region mass surrounded by an vacuous shell, or
``halo''.\footnote{In reality, the dust approximation is never exact,
  and with realistic quantum fields it is impossible to suppress the
  propagation of signals out from the perturbed region entirely.  Here
  we assume that the dust approximation is good and that the
  corrections (the weak signals leaking out) can be precisely
  cancelled by subtle perturbations of the global background, without
  affecting the basic geometry and coarse-grained entropy of our
  solutions.}

\begin{figure}[!htb] \subfigure[]{ \includegraphics [scale=.7]
  {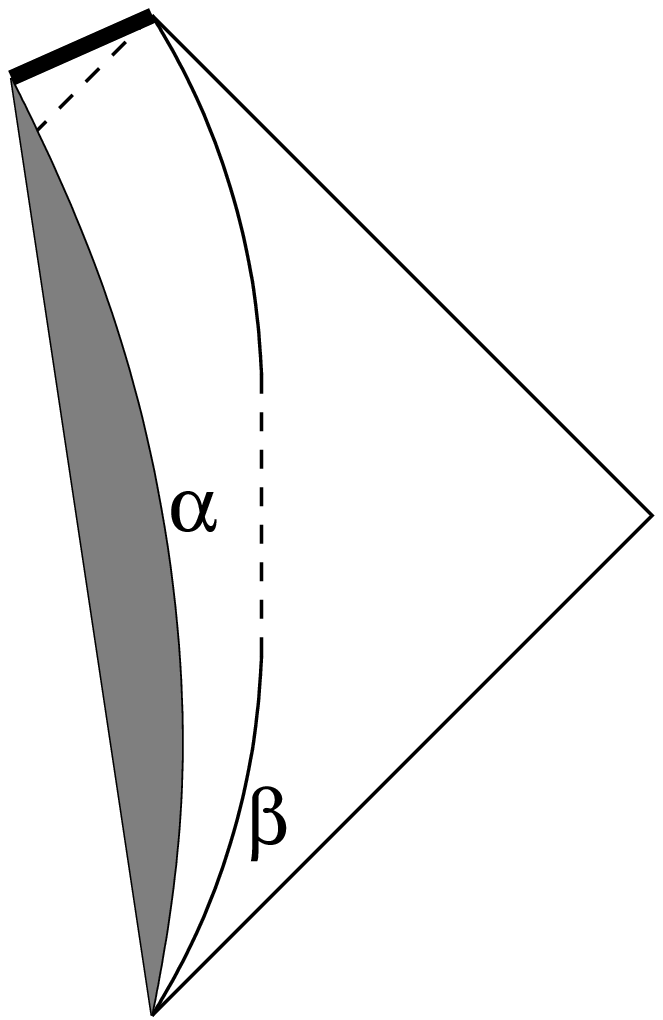} } \hspace{0.3 in} \subfigure []{\includegraphics [scale =
    .7]{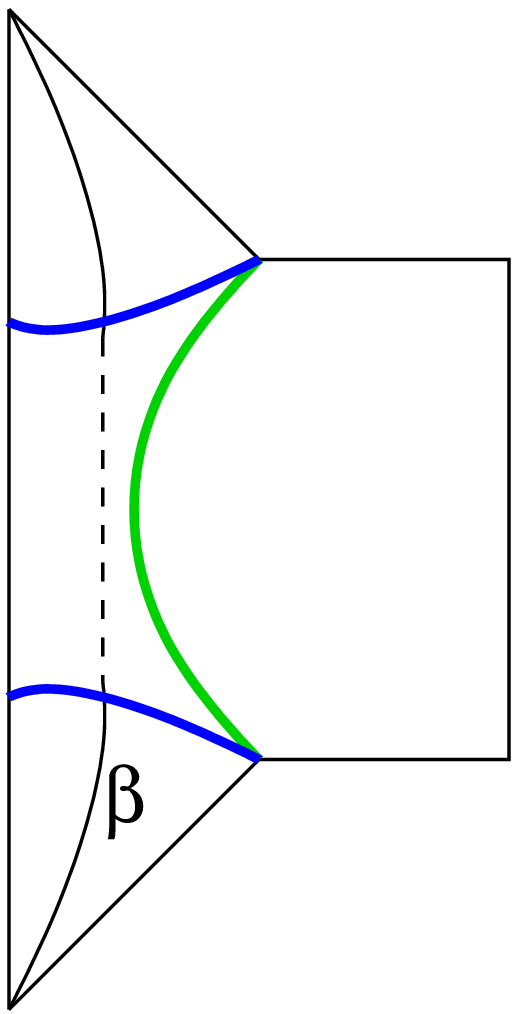}} \hspace{.3 in} \subfigure []{\includegraphics
  [scale=.7] {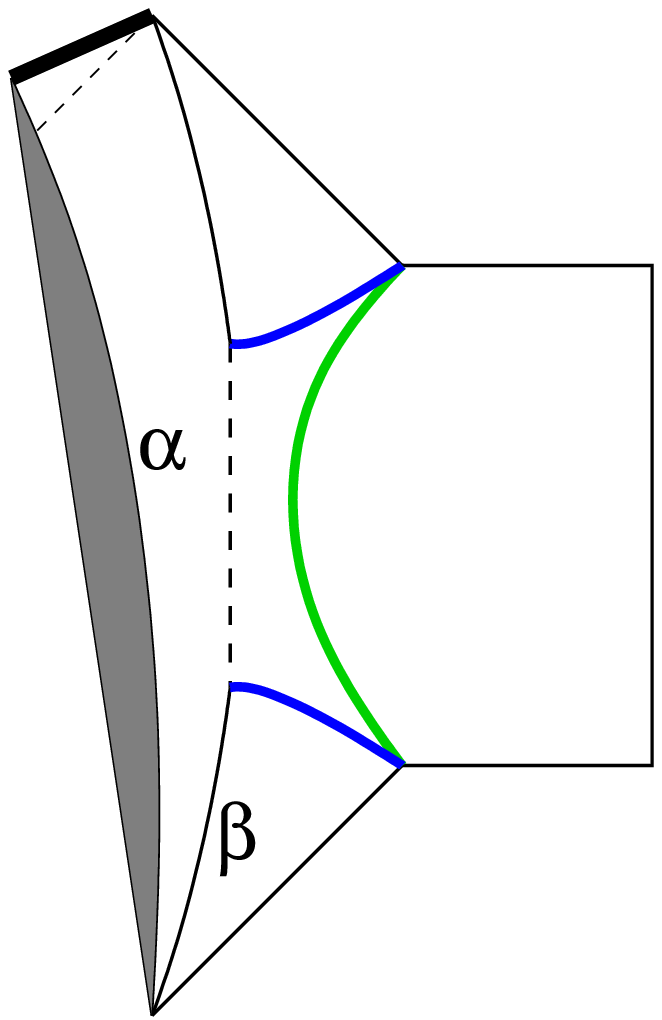}} \caption{Construction of an in-state
  which will {\em not} cause a global collapse. The full solution (c)
  is obtained by gluing a piece of the background solution (b) to a
  collapsing ball of dust surrounded by vacuum (a). The full solution
  keeps the region of (a) to the left of the line $\beta$ and the
  region of (b) to the right of $\beta$. $\alpha$ marks the boundary
  between the collapsing dust (shaded) and vacuum. In the center of
  the diagram the particles of the outside solution have been
  absorbed, rendering the choice of $\beta$ arbitrary (dashed
  line). In this region $\beta$ is not a geodesic, whereas $\alpha$ is
  a geodesic throughout the diagram.  Heavy lines represent the domain
  wall (timelike) and the reheating/recooling surfaces (spacelike).
  The black hole horizon and singularity are shown in the top left
  corner.}  \label{halo} \end{figure}

Our solution consists of three pieces joined along timelike
hypersurfaces, as shown in Fig.~\ref{halo}.  We take the outer piece
to be a portion of the background, i.e., an open contracting dusty FRW
universe, except for the region $\rho<\rho_0$, which we excise and
throw away.  [Here $\rho$ is the comoving radial coordinate; see
  Eq.~(\ref{eq-iop}).]  By starting with this exterior piece and
working our way in, the absence of any perturbation outside $\rho_0$
is guaranteed by construction. This will be important.

We take the middle piece to be vacuous.  Then Birkhoff's theorem
implies that the middle piece is given by a portion of a Schwarzschild
solution.  The Schwarzschild radius characterizing this solution is
that of the black hole which the excised piece (which we discarded)
would have formed if complemented by a vacuum exterior.  Equivalently,
the correct solution is fixed by the requirement that the
Schwarzschild geometry contain a geodesic identical to the path of the
innermost dust shell of the outer piece.  This geodesic, $\beta$,
forms the outer limit of the middle piece.

We have thus obtained an open universe containing a black hole.  The
fully extended solution contains an Einstein-Rosen bridge to an
asymptotically flat region.  This would not describe a reasonable
perturbation of the in-state.  But this is easily fixed by bringing
matter back at a smaller radius.  

Consider a timelike radial geodesic, $\alpha$, in the middle piece,
such that $\alpha$ begins at past timelike infinity and lies
everywhere inside $\beta$, as shown in Fig.~\ref{halo}.  The inner
piece (inside of $\alpha$) can then be taken to be the central portion
of an open dust-dominated FRW universe uniquely determined by the
requirement that it contain the geodesic hypersurface $\alpha$ that
forms its boundary.  Note that the inner piece is {\em not\/} a
portion of the original open FRW universe---it will have a different
density at the time when curvature starts to dominate.\footnote{One
  could also construct halo solutions with white-hole initial
  conditions by taking the inner piece to be a portion of a closed
  recollapsing dust dominated FRW universe.}

Because they involve underdense shells, we will call these states halo
states.  They differ from the background only within a finite comoving
radius, and because they contain only dust, no signal will propagate
out to the unperturbed region.  Hence, it is easy to evolve the halo
states from past infinity to the recooling surface.  In the
unperturbed region, the conversion of particles into vacuum energy
will proceed exactly as in the vacuum.\footnote{At the recooling
  surface, spatial gradients will arise in the field which makes up
  the domain wall, allowing signals to propagate out.  This is
  acceptable, since our goal was only to avoid disturbing the
  recooling.  However, there will be corrections to our solution after
  recooling.}

The halo region will not be absorbed, so that a collapsing ball of
dust will enter the central region of the spacetime.  Thus halo states
form black holes, and black holes cannot fit if they are too large.
It follows that we can estimate the number of (bouncing) halo states
by the entropy of the black holes they form, which is of order
$R^2/G$.  Thus we find that the entropy bound (\ref{eq-bound}) is
approximately saturated by the non-crunching states.

Of course, the idealization as dust is not realistic in quantum field
theory.  Even if the background outside $\rho_0$ is truly unperturbed
near ${\cal I}^-$, some weak signals will propagate out from the halo
and disturb the absorption process on large scales.  However, we do
expect that there are allowed perturbations whose main feature is a
halo of size $\rho_0$, plus extremely subtle additional perturbations
outside the halo which cancel against such signals.

Not all allowed states will be of halo form.  We have merely pointed
out that the halo states appear to be the only allowed states that can
be described simply near the boundary of spacetime, and that they can
account for a significant portion of the allowed entropy.  The halo
states are highly inhomogeneous and thus of no relevance to a
realistic universe. The other allowed states are extremely scrambled
near infinity and hence hard to characterize.  

As we discussed in the introduction, this limits the appeal of an
S-matrix description (involving the past hat) as a framework for the
landscape.  But the attractive properties of the Coleman-De~Luccia
solution for defining observables~\cite{FreSus04}---the presence of
noninteracting particles and low curvature at late times---remain to
be exploited.  They may yet allow for the construction of an exact
quantum cosmology whose amplitudes correspond to asymptotic
observables in the future hat.

\subsubsection*{Acknowledgements}

We would like to thank T.~Banks and L.~Susskind for valuable
discussions.  This work was supported by the Berkeley Center for
Theoretical Physics, by a CAREER grant of the National Science
Foundation, and by DOE grant DE-AC03-76SF00098.

\bibliographystyle{board}
\bibliography{all}

\end{document}